\documentclass[twocolumn,superscriptaddress]{revtex4}%
\usepackage{graphicx}
\usepackage{dcolumn}
\usepackage{bm}
\usepackage{color}
\usepackage{amsmath}
\usepackage{amsfonts}
\usepackage{amssymb}
\usepackage{ulem}%

\begin{document}

\title{Intermediate-state Coulomb-corrected strong-field approximation for rescattering processes}
\author{Chunli Miao}
\affiliation{Institute of Theoretical Physics, State Key Laboratory of Quantum Optics Technologies and Devices, Shanxi University, Taiyuan 030006, China}
\author{Jiarui Qin}
\affiliation{Institute of Theoretical Physics, State Key Laboratory of Quantum Optics Technologies and Devices, Shanxi University, Taiyuan 030006, China}
\author{Chan Li}
\email{lichan@sxu.edu.cn}
\affiliation{Institute of Theoretical Physics, State Key Laboratory of Quantum Optics Technologies and Devices, Shanxi University, Taiyuan 030006, China}
\author{Xiaolei Hao}
\email{xlhao@sxu.edu.cn}
\affiliation{Institute of Theoretical Physics, State Key Laboratory of Quantum Optics Technologies and Devices, Shanxi University, Taiyuan 030006, China}
\author{Weidong Li}
\affiliation{Shenzhen Key Laboratory of University Laser and Advanced Material Technology, Shenzhen Technology University, Shenzhen 518118, China}
\author{Jing Chen}
\email{chenjing@ustc.edu.cn}
\affiliation{Department of Modern Physics, and Hefei National Research Center for Physical
Sciences at the Microscale and School of Physical Sciences, University of
Science and Technology of China, Hefei 230026, China}
\affiliation{Hefei National Laboratory, Hefei 230088, China}

\begin{abstract}
We analytically derive the all-order strong-field S-matrix series
incorporating intermediate-state Coulomb-Volkov corrections (ICSFA). Focusing
on rescattering processes described by the second-order term, we
systematically investigate the impact of intermediate-state Coulomb
interactions on above-threshold ionization (ATI) spectra of atomic hydrogen in
linearly polarized laser fields. Crucially, ICSFA spectra demonstrate superior
agreement with the results obtained by numerically solving the time-dependent
Schr\"{o}dinger equation compared to the standard strong-field approximation
(SFA) and final-state Coulomb-corrected SFA (FCSFA). Our analysis reveals that
intermediate-state Coulomb corrections enhance the yield of the third- and
fourth-return-recollision trajectories while modifying interference patterns
in the energy spectrum. The observed enhancement of the
multi-return-recollision trajectories can be attributed to modifications of
the ionization yield and scattering cross-section, which are induced by
intermediate-state Coulomb effects. These effects are equivalent to the
so-called Coulomb focusing effect.

keywords: rescattering; the Coulomb interactions; above-threshold ionization; the time-dependent
Schr\"{o}dinger equation; multi-return-recollision trajectories; scattering cross-section; Coulomb focusing effect
\end{abstract}

\maketitle

\bigskip

%%%%%%%%%%%%%%%%%%%%%%%%%%  body  %%%%%%%%%%%%%%%%%%%%%%%%%%

\section{INTRODUCTION}

In ultrafast physics, photoionization processes have provided an
unprecedented insight into the inner working of atoms and molecules.
As a fundamental physical process, it serves as the underlying basis
for a series of key phenomena, including high-harmonic generation (HHG)
\cite{Krausz2009}
which is driving substantial research interest as a promising platform for
extreme ultraviolet sources, ultrafast micro-dynamics, and biomedical innovations
\cite{Ferray1988,Guo2015,Yu2015,Gohle2005,Bai2021,Heide2022,Lv2021,Cui2025},
among others. The strong-field approximation (SFA) is a widely and successfully
used theoretical method to study these processes. Keldysh initially applied this
theory in the form of the so-called Keldysh-Faisal-Reiss theory to study the photoionization
of atoms and molecules in laser fields \cite{Keldysh1965,Faisal1973,Reiss1980}.
Since then, various versions of SFA have been developed to understand the
physics of matter in intense laser fields \cite{Amini2019}. It is also well
known that the original SFA, based on the plane-wave Volkov state, does not
account for the Coulomb interaction of the electron with the parent ion.

Plenty of efforts have been made \cite{Popruzhenko2014} in order to include
the Coulomb effect to the ionization process in SFA. These corrections can be
divided into two main categories: one is the modification of the state, that
is, replacing the final state of Volkov state with the Coulomb-Volkov state,
which is called the Coulomb-Volkov approximation (CVA)
\cite{Jingtao2007,Milo1998,Faisal2016,Faisal2018,Milo2019,Leone1987}, and the
other is the orbit-based Coulomb correction by including the influence of
Coulomb potential on the action and trajectories
\cite{TM2012,Popv2008,Popruzh2008,TM2010,Hao2022,lai2015}.

A number of strong-field phenomena, particularly in ionization experiments,
demonstrate features caused by the ion's Coulomb potential that evade the SFA.
These include low-energy structures (LES)
\cite{Rudenko2005,Blaga2009,Quan2009,Wu2012,Quan2016,Shafir2013,Li2013,Liu2010,Guo2013,Liu2021}
and the significant role of multiple-return-recollision (MRR) trajectories in
the recollision process. The Coulomb focusing effect has been demonstrated to
significantly enhance the contribution of MRR trajectories
\cite{Brabec1996,Yudin2001}. In certain scenarios, these contributions may
even surpass those of the first return orbit \cite{XM2003,Hao2011,XY2013}.
Recently, a joint theoretical and experimental investigation indicated that
the Coulomb field can increase the contribution of the MRR trajectories and
the cutoffs in the high-order above-threshold ionization of Ar atom
\cite{HaoPRA2020}. In these phenomena, the Coulomb effect on the intermediate
rescattering process plays an essential role. However, so far, the CVA for SFA
is limited to corrections to the final state
\cite{Faisal2016,Faisal2018} which does not account for the above phenomena.

In this paper, we derive the complete S-matrix series for Coulomb-Volkov
correction to the intermediate state in SFA (ICSFA).
We find that the
intermediate-state Coulomb interaction can effectively improve the
contribution of MRR trajectories, which is in agreement with the conclusion
obtained in Ref. \cite{HaoPRA2020}. This intermediate-state Coulomb effect is anticipated to manifest
significantly in other strong-field recollision-based processes, such as HHG for
attosecond pulse production. It should also be explicitly accounted for in ultrafast
imaging techniques, including laser-induced electron diffraction
\cite{Meckel2008,Blaga2012,Xu2014,Pullen2016,Pullen2015,Wolter2016}. The paper is arranged as follows. To start
with, we give the details to obtain the complete ICSFA in all orders. Next, we
conduct rigorous comparisons of the photoelectron
momentum distributions (PMDs) and the photoelectron energy spectra for hydrogen
atoms in linearly polarized intense laser fields, contrasting results from
SFA, ICSFA, final-state Coulomb-corrected SFA (FCSFA), and time-dependent
Schr\"{o}dinger equation (TDSE) simulations. Then, we perform a detailed
analysis on the Coulomb effects to the intermediate-state in the rescattering
process. Finally, we draw a conclusion.

\section{METHODS}

The Hamiltonian $H(t)\ $of an atom in an intense laser field is (in atomic
units, $m=|e|=\hbar=1$)%
\begin{equation}
H(t)=H_{0}+V_{I}(t), \label{HI}%
\end{equation}
where $H_{0}$ is the atomic unperturbed Hamiltonian
\begin{equation}
H_{0}=\frac{\widehat{\mathbf{p}}^{2}}{2}+V(r),
\end{equation}
with $\widehat{\mathbf{p}}=-i\nabla$ as the momentum operator, and $V(r)$ representing the potential experienced by the
electron. The term
$V_{I}$ in Eq. (\ref{HI}) corresponds to the interaction with the laser field (in velocity gauge and dipole approximation)
\begin{equation}
V_{I}=\mathbf{A}\left(  t\right)  \cdot\widehat{\mathbf{p}}+\frac
{\mathbf{A}^{2}(t)}{2}, \label{VI}%
\end{equation}
 where $\mathbf{A}\left(  t\right)
$ is the vector potential of the laser field.

In the spirit of SFA, the total Hamiltonian can be divided in different ways.
Eq. (\ref{HI}) is a way of dividing. Usually, the Volkov-state partition of
the total Hamiltonian is also used:%
\begin{equation}
H(t)=H_{V}(t)+V_{V}(t), \label{HV}%
\end{equation}
with
\begin{equation}
H_{V}(t)=\frac{\left[  \mathbf{p}+\mathbf{A}(t)\right]  ^{2}}{2},
\end{equation}
and
\begin{equation}
V_{V}(t)=V(r). \label{equ:v0}%
\end{equation}

Then, the total Green's function corresponding to $H(t)$ can be expanded as
\cite{Becker2005}
\begin{align}
G\left(  t,t^{\prime}\right)  =  &  G_{V}\left(  t,t^{\prime}\right)
\nonumber\\
&  +\int_{t^{\prime}}^{t}dt_{1}G_{V}\left(  t,t_{1}\right)  V_{V}\left(
t_{1}\right)  G_{V}\left(  t_{1},t^{\prime}\right) \nonumber\\
&  +\int_{t^{\prime}}^{t}\int_{t_{1}}^{t}dt_{2}dt_{1}G_{V}\left(
t,t_{2}\right)  V_{V}\left(  t_{2}\right) \nonumber\\
&  \times G_{V}\left(  t_{2},t_{1}\right)  V_{V}\left(  t_{1}\right)
G_{V}\left(  t_{1},t^{\prime}\right) \nonumber\\
&  +\cdot\cdot\cdot, \label{GV}%
\end{align}
where%
\begin{equation}
G_{V}(t,t^{\prime})=-i\theta(t-t^{\prime})\sum_{p}\left\vert \Phi_{\mathbf{p}%
}^{V}(\mathbf{r},t)\right\rangle \left\langle \Phi_{\mathbf{p}}^{V}%
(\mathbf{r},t^{\prime})\right\vert , \label{equ:Gv}%
\end{equation}
corresponds to $H_{V}(t)$, with $\Phi_{\mathbf{p}}^{V}(\mathbf{r},t)$
representing the Volkov state with momentum $\mathbf{p}$: $\Phi_{\mathbf{p}%
}^{V}(\mathbf{r},t) =\frac{1}{(2\pi)^{3/2}}e^{i\mathbf{p}\cdot\mathbf{r}}
e^{-i\int^{t}\frac{[\mathbf{p}+\mathbf{A}(\tau)]^{2}}{2}d\tau}$, and
$\theta(t-t^{\prime})$ denoting the Heaviside function.

Faisal \cite{Faisal2016,Faisal2018} has developed an S-matrix theory in which
final-state Coulomb interactions are taken into account in all orders. He
determined the exact Coulomb-Volkov Hamiltonian, and then constructed the
exact Coulomb-Volkov propagator in order to identify the rest interaction
associated with the Coulomb-Volkov state. Here we will follow his idea to
derive the S-matrix theory in which intermediate-state Coulomb interactions
are taken into account in all orders. To include the effect of electron-ion
Coulomb interaction, the Coulomb-Volkov-state partition of the total
Hamiltonian is introduced:
\begin{equation}
H(t)=H_{CV}(t)+V_{CV}(t),
\end{equation}
where $H_{CV}(t)$ is the Coulomb-Volkov-state reference Hamiltonian and
$V_{CV}(t)$ is the Coulomb-Volkov-state rest interaction. $H_{CV}(t)$ has the
following formula \cite{Faisal2016,Faisal2018}:%
\begin{equation}
H_{CV}(t)=\frac{\widehat{\mathbf{p}}^{2}}{2}-\frac{Z}{r}+\frac{\mathbf{A}%
^{2}(t)}{2}+\mathbf{A}(t)\cdot\widehat{\mathbf{Q}}(t), \label{Hcv}
\end{equation}
with $Z$ the nuclear charge and the vector operator
\begin{equation}
\widehat{\mathbf{Q}}(t)\equiv\underset{\mathbf{p}}{\sum}|\phi_{\mathbf{p}}%
^{C}\rangle\mathbf{p}\langle\phi_{\mathbf{p}}^{C}|.
\end{equation}
In the above equation, $\langle\mathbf{r}|\phi_{\mathbf{p}}^{C}\rangle$ stands
for the well-known `ingoing' stationary Coulomb waves $\phi_{\mathbf{p}}%
^{C}(\mathbf{r})$ with momentum $\mathbf{p}$ \cite{Landau1965}:%
\begin{align}
\phi_{\mathbf{p}}^{C}(\mathbf{r})=  &  (2\pi)^{-3/2}e^{i\mathbf{p}%
\cdot\mathbf{r}}{}N_{p}\nonumber\\
&  \times_{1}F_{1}(-i\eta_{p},1,-i(pr+\mathbf{p}\cdot\mathbf{r})),
\label{equ:cv}%
\end{align}
where $_{1}F_{1}(-i\eta_{p},1,-i(pr+\mathbf{p}\cdot\mathbf{r}))$ is the
confluent hypergeometric function, and
\begin{equation}
N_{p}=\exp\left(  (\pi/2)\eta_{p}\right)  \Gamma(1+i\eta_{p}), \label{Np}%
\end{equation}
is the normalization coefficient, with $\eta_{p}=\frac{1}{p},p=|\mathbf{p}|$,
and $\Gamma(1+i\eta_{p})$ is the gamma function.

Notably, the term $\mathbf{A}(t)\cdot\widehat{\mathbf{Q}}(t)$ in
Eq. (\ref{Hcv}) represents the coupling between the laser field and the
Coulomb potential of the atom. The inclusion of this term ensures that
the wave function of Coulomb-Volkov state becomes an exact
eigenstate of $H_{CV}(t)$. Then the Coulomb-Volkov state rest interaction $V_{CV}(t)$\ can be obtained as%
\begin{equation}
V_{CV}(t)\equiv H(t)-H_{CV}(t)=V(r)+\frac{Z}{r}+\mathbf{A}\left(
t\right) \cdot\left[
\widehat{\mathbf{p}}-\widehat{\mathbf{Q}}(t)\right]  .
\end{equation}
The total Green's function can also be expanded based on the Coulomb-Volkov
state \cite{Faisal2016}:%
\begin{align}
G\left(  t,t^{\prime}\right)  =  &  G_{CV}\left(  t,t^{\prime}\right)
\nonumber\\
&  +\int_{t^{\prime}}^{t}dt_{1}G_{CV}\left(  t,t_{1}\right)  V_{CV}\left(
t_{1}\right)  G_{CV}\left(  t_{1},t^{\prime}\right) \nonumber\\
&  +\int_{t^{\prime}}^{t}\int_{t_{1}}^{t}dt_{2}dt_{1}G_{CV}\left(
t,t_{2}\right)  V_{CV}\left(  t_{2}\right) \nonumber\\
&  \times G_{CV}\left(  t_{2},t_{1}\right)  V_{CV}\left(  t_{1}\right)
G_{CV}\left(  t_{1},t^{\prime}\right) \nonumber\\
&  +\cdot\cdot\cdot, \label{GCV}%
\end{align}
where%
\begin{equation}
G_{CV}(t,t^{\prime})=-i\theta(t-t^{\prime})\sum_{p}\left\vert \Phi
_{\mathbf{p}}^{CV}(\mathbf{r},t)\right\rangle \left\langle \Phi_{\mathbf{p}%
}^{CV}(\mathbf{r},t^{\prime})\right\vert ,
\end{equation}
with $\Phi_{\mathbf{p}}^{CV}(\mathbf{r},t)$ the Coulomb-Volkov wave function:%
\begin{equation}
\Phi_{\mathbf{p}}^{CV}(\mathbf{r},t)=\phi_{\mathbf{p}}^{C}(\mathbf{r}%
)e^{-i\int^{t}\frac{[\mathbf{p}+\mathbf{A}(\tau)]^{2}}{2}d\tau}.
\end{equation}

A generally useful expression for the solution of any Schr\"{o}dinger equation
satisfying a given initial and/or a final state condition is given by the
following closed form \cite{Becker2005}%
\begin{align}
\left\vert \Psi\left(  t\right)  \right\rangle =  &  \left\vert \phi
_{i}\left(  t\right)  \right\rangle \nonumber\\
&  +\int^{t}dt_{1}G_{f}\left(  t,t_{1}\right)  V_{I}\left(  t_{1}\right)
\left\vert \phi_{i}\left(  t_{1}\right)  \right\rangle \nonumber\\
&  +\int^{t}\int_{t_{1}}^{t}dt_{2}dt_{1}G_{f}\left(  t,t_{2}\right)
V_{f}\left(  t_{2}\right) \nonumber\\
&  \times G\left(  t_{2},t_{1}\right)  V_{I}\left(  t_{1}\right)  \left\vert
\phi_{i}\left(  t_{1}\right)  \right\rangle , \label{Wvt}%
\end{align}
where $\left\vert \phi_{i}\left(  t\right)  \right\rangle $ is the initial
state, $G_{f}\left(  t,t_{1}\right)  $ is the final state reference Green's
function corresponding to the Hamiltonian $H_{f}(t)$, $V_{f}(t)\equiv
H(t)-H_{f}(t)$ is the associated final state rest-interaction. Projecting the
wave function in Eq. (\ref{Wvt}) onto the final state $\langle\phi_{f}\left(
t\right)  |$ and letting $t\rightarrow\infty$, the transition amplitude can be
obtained as%
\begin{align}
S_{fi}=  &  \langle\phi_{f}\left(  t\right)  \left\vert \phi_{i}\left(
t\right)  \right\rangle \nonumber\\
&  -i%
%TCIMACRO{\dint }%
%BeginExpansion
{\displaystyle\int}
%EndExpansion
dt_{1}\left\langle \phi_{f}\left(  t_{1}\right)  \right\vert V_{I}\left(
t_{1}\right)  \left\vert \phi_{i}\left(  t_{1}\right)  \right\rangle
\nonumber\\
&  -i\int\int_{t_{1}}dt_{2}dt_{1}\left\langle \phi_{f}\left(  t_{2}\right)
\right\vert V_{f}\left(  t_{2}\right) \nonumber\\
&  \times G\left(  t_{2},t_{1}\right)  V_{I}\left(  t_{1}\right)  \left\vert
\phi_{i}\left(  t_{1}\right)  \right\rangle . \label{Sfi}%
\end{align}
After expanding the total Green's function in the second term, one can obtain
all orders of the S-matrix series.

In Ref. \cite{Faisal2016}, the Coulomb-Volkov state was selected as the final
reference state in Eq. (\ref{Wvt}), i.e., $G_{f}\left(  t,t_{1}\right)  =$
$G_{CV}\left(  t,t_{1}\right)  $, and the total Green's function $G\left(
t_{2},t_{1}\right)  $ in Eq. (\ref{Sfi}) is expanded based on the Volkov state
by using Eq. (\ref{GV}). The S-matrix series for FCSFA then can be obtained:%
\begin{equation}
S_{fi}=\underset{n=1}{\overset{\infty}{\sum}}S_{fi}^{(n)},
\end{equation}
with%
\begin{align}
S_{fi}^{F(1)}=  &  -i\int dt_{1}\left\langle \Phi_{\mathbf{p}}^{CV}\left(
\mathbf{r}_{1},t_{1}\right)  \right\vert \nonumber\\
&  \times\left[  \mathbf{A}\left(  t_{1}\right)  \cdot\widehat{\mathbf{p}}
+\frac{\mathbf{A}^{2}(t_{1})}{2}\right]  \left\vert \phi_{i}\left(
\mathbf{r}_{1},t_{1}\right)  \right\rangle , \label{SF1}%
\end{align}%
\begin{align}
S_{fi}^{F(2)}=  &  -i\int\int_{t_{1}}dt_{2}dt_{1}\left\langle \Phi
_{\mathbf{p}}^{CV}\left(  \mathbf{r}_{2},t_{2}\right)  \right\vert \nonumber\\
&  \times  \left\{V(r_{2})+\frac{Z}{r_{2}}+\mathbf{A}(t_{2})\cdot\left[\widehat{\mathbf{p}}-\widehat
{\mathbf{Q}}(t_{2})\right]\right\} \nonumber\\
&  \times G_{V}\left(  \mathbf{r}_{2},t_{2};\mathbf{r}_{1},t_{1}\right)
\nonumber\\
&  \times\left[  \mathbf{A}\left(  t_{1}\right)  \cdot\widehat{\mathbf{p}}
+\frac{\mathbf{A}^{2}(t_{1})}{2}\right]  \left\vert \phi_{i}\left(
\mathbf{r}_{1},t_{1}\right)  \right\rangle , \label{SF2}%
\end{align}
and the general $n$th-order amplitude%
\begin{align}
S_{fi}^{F(n)}=  &  -i\int\cdot\cdot\cdot\int_{t_{n-2}}\int_{t_{n-1}}%
dt_{n}dt_{n-1}\cdot\cdot\cdot dt_{1}\langle\Phi_{\mathbf{p}}^{CV}(\mathbf{r}_{n},t_{n})|\nonumber\\
&  \times\left\{V(r_{n})+\frac{Z}{r_{n}}
 +\mathbf{A}(t_{n})\cdot\left[\widehat{\mathbf{p}} -\widehat{\mathbf{Q}}
(t_{n})\right]\right\} \nonumber\\
&  \times G_{V}(\mathbf{r}_{n},t_{n};\mathbf{r}_{n-1},t_{n-1})V(r_{n-1})\times\cdot\cdot\cdot\times \nonumber\\
%&\nonumber\\
&  \times G_{V}(\mathbf{r}_{2},t_{2};\mathbf{r}_{1},t_{1})\left[
\mathbf{A}\left(  t_{1}\right)  \cdot\widehat{\mathbf{p}} +\frac
{\mathbf{A}^{2}(t_{1})}{2}\right] \nonumber\\
&  \times|\phi_{i}(\mathbf{r}_{1},t_{1})\rangle. \label{SFn}%
\end{align}

In this work, we select the Volkov state as the final reference state in Eq.
(\ref{Wvt}), i.e., $G_{f}\left(  t,t_{1}\right)  =$ $G_{V}\left(
t,t_{1}\right)  $, but the further expansion to the total Green's function is
based on the Coulomb-Volkov state by using Eq. (\ref{GCV}). Consequently, we
arrive at the desired all-order ICSFA S-matrix series with%
\begin{align}
S_{fi}^{I(1)}=  &  -i\int dt_{1}\langle\Phi_{\mathbf{p}}^{V}\left(
\mathbf{r}_{1},t_{1}\right)  |\nonumber\\
&  \times\left[  \mathbf{A}\left(  t_{1}\right)  \cdot\widehat{\mathbf{p}}
+\frac{\mathbf{A}^{2}(t_{1})}{2}\right]  \left\vert \phi_{i}\left(
\mathbf{r}_{1},t_{1}\right)  \right\rangle , \label{SI1}%
\end{align}%
\begin{align}
S_{fi}^{I(2)}=  &  -i\int\int_{t_{1}}dt_{2}dt_{1}\left\langle \Phi
_{\mathbf{p}}^{V}\left(  \mathbf{r}_{2},t_{2}\right)  \right\vert  \nonumber\\
&  \times V(r_{2})G_{CV}\left(  \mathbf{r}_{2},t_{2};\mathbf{r}_{1},t_{1}\right)
\nonumber\\
&  \times\left[  \mathbf{A}\left(  t_{1}\right)  \cdot\widehat{\mathbf{p}}
+\frac{\mathbf{A}^{2}(t_{1})}{2}\right]  \left\vert \phi_{i}\left(
\mathbf{r}_{1},t_{1}\right)  \right\rangle , \label{SI2}%
\end{align}
and the nth-order term%
\begin{align}
S_{fi}^{I(n)}=  &  -i\int\cdot\cdot\cdot\int_{t_{n-2}}\int_{t_{n-1}}%
dt_{n}dt_{n-1}\cdot\cdot\cdot dt_{1}\nonumber\\
&  \times\langle\Phi_{\mathbf{p}}^{V}(\mathbf{r}_{n},t_{n})|V(r_{n})
 G_{CV}(\mathbf{r}_{n},t_{n};\mathbf{r}_{n-1},t_{n-1})\nonumber\\
&  \times  \left\{V(r_{n-1})+\frac{Z}{r_{n-1}}+\mathbf{A}(t_{n-1})\cdot\left[\widehat{\mathbf{p}}-\widehat
{\mathbf{Q}}(t_{n-1})\right]\right\} \nonumber\\ \nonumber\\
&  \times\cdot\cdot\cdot\times\nonumber\\
&  \times G_{CV}(\mathbf{r}_{2},t;\mathbf{r}_{1},t_{1})\left[  \mathbf{A}%
(t_{1})\cdot\widehat{\mathbf{p}}+\frac{\mathbf{A}^{2}(t_{1})}{2}\right]
\nonumber\\
&  \times|\phi_{i}(\mathbf{r}_{1},t_{1})\rangle. \label{SIn}%
\end{align}

The first-order term of ICSFA $S_{fi}^{I(1)}$ in Eq. (\ref{SI1}) is the same
as that in the traditional SFA, describing the direct ionization process in
which the electron is ionized by the laser field from the ground state to a
final Volkov state. Although the first-order term of FCSFA $S_{fi}^{F(1)}$ in
Eq. (\ref{SF1}) also depicts the direction ionization process, its final state
is replaced by the Coulomb-Volkov state. The second-order term of FCSFA
$S_{fi}^{F(2)}$ in Eq. (\ref{SF2}) describes the process in which the electron
is firstly ionized to the Volkov state and then experiences the interaction
$V_{CV}(t)$ to become a Coulomb-Volkov state. This does not directly
correspond to the ordinary rescattering term in which the interaction in the
second step is the ionic potential $V_{V}(t)=V(r)$. Whereas the second-order
term of ICSFA $S_{fi}^{I(2)}$ in Eq. (\ref{SI2}) has a clear meaning of the
rescattering process. Its difference from the ordinary rescattering term lies
in the intermediate state, which is a Coulomb-Volkov state instead of a Volkov
state. This modification will incorporate the Coulomb effect on the
intermediate state.

It is widely recognized that higher-order expansions pose
fundamental challenges for the SFA framework. For both the
conventional SFA and FCSFA, the convergence of the series remains
unproven and constitutes an open mathematical problem
\cite{Faisal2016}. ICSFA, which also builds on the SFA framework,
inherently inherits this deep theoretical challenge. Consequently,
the practical validity of truncating the series at any finite
order (whether for SFA, FCSFA, or ICSFA) cannot be theoretically
guaranteed; it must instead be validated empirically through
case-by-case comparisons with experimental data or higher-order
benchmarks. Despite these limitations, here we compare the three
theories at the same order, specifically the second order, which
is widely recognized as dominating the rescattering process to
isolate how intermediate-state Coulomb interactions modify the
rescattering process.

The second-order term of the S-matrix series can be rewritten in a more
concrete term as
\begin{equation}
M_{resc}(\mathbf{p})=-\int_{-\infty}^{\infty}\int_{t_{1}}^{\infty}dt_{2}%
dt_{1}\int d^{3}\mathbf{k}e^{iS_{\mathbf{p}}(t_{1},t_{2},\mathbf{k}%
)}V_{\mathbf{pk}}V_{\mathbf{k}g},\label{M}%
\end{equation}
with the classical action%
\begin{align}
S_{\mathbf{p}}(t_{1},t_{2},\mathbf{k})= &  \frac{1}{2}\int_{-\infty}^{t_{2}%
}d\tau\left[  \mathbf{p}+\mathbf{A}(\tau)\right]  ^{2}\nonumber\\
&  -\frac{1}{2}\int_{t_{1}}^{t_{2}}d\tau\lbrack\mathbf{k}+\mathbf{A}%
(\tau)]^{2}+I_{p}t_{1},\label{S}%
\end{align}
where $\mathbf{k}$ and $\mathbf{p}$ are the canonical momentum before and
after scattering, respectively; and $I_{p}$ is the ionization potential. The
electron, initially in the ground state $|\phi_{i}\left(  t_{1}\right)
\rangle$, is ionized into its intermediate state at the time $t_{1}$, then
rescattered off the binding potential at the time $t_{2}$. The classical
action is the same in the three forms of SFA, ICSFA, and FCSFA, but the
prefactors are different. In the SFA framework, the second-order term is
conventionally referred to as the improved SFA (ISFA). In this work, our use of SFA
universally encompasses all orders of the S-matrix series, with no Coulomb
corrections applied to either final or intermediate states.

Here, we adopt the ionization of the hydrogen atom from
its ground $1s$ state ($Z$ = 1) as a clean theoretical benchmark,
where the potential is given by $V(r)=-1/r$. The prefactors for
SFA in this case are
\begin{subequations}
\begin{align}
V_{\mathbf{k}g} &  =\langle\Phi_{\mathbf{k}}^{V}(\mathbf{r})|\left[
\mathbf{A}\left(  t_{1}\right)  \cdot\widehat{\mathbf{p}}+\frac{\mathbf{A}%
^{2}(t_{1})}{2}\right]  |\phi_{i}(\mathbf{r})\rangle\nonumber\label{vkg}\\
&  =\frac{4}{\sqrt{2}\pi}\left[  \mathbf{A}\left(  t_{1}\right)
\cdot\mathbf{k}+\frac{\mathbf{A}^{2}(t_{1})}{2}\right]  \frac{1}{\left(
1+\mathbf{k}^{2}\right)  ^{2}},\\
V_{\mathbf{pk}} &  =\langle\Phi_{\mathbf{p}}^{V}\left(  \mathbf{r}\right)
|-\frac{1}{r}|\Phi_{\mathbf{k}}^{V}(\mathbf{r})\rangle\nonumber\\
&  \overset{\mathbf{q}=\mathbf{k-p}}{=}-\frac{1}{2\pi^{2}}\frac{1}%
{\mathbf{q}^{2}}.
\end{align}
In FCSFA, they are \cite{Faisal2016,Faisal2018}
\end{subequations}
\begin{subequations}
\begin{align}
V_{\mathbf{k}g} &  =\langle\Phi_{\mathbf{k}}^{V}(\mathbf{r})|\left[
\mathbf{A}\left(  t_{1}\right)  \cdot\widehat{\mathbf{p}}+\frac{\mathbf{A}%
^{2}(t_{1})}{2}\right]  |\phi_{i}(\mathbf{r})\rangle\nonumber\label{vkg-F}\\
&  =\frac{4}{\sqrt{2}\pi}\left[  \mathbf{A}\left(  t_{1}\right)
\cdot\mathbf{k}+\frac{\mathbf{A}^{2}(t_{1})}{2}\right]  \frac{1}{\left(
1+\mathbf{k}^{2}\right)  ^{2}},\\
V_{\mathbf{pk}} &  =\langle\Phi_{\mathbf{p}}^{CV}\left(  \mathbf{r}\right)
|\left[  \mathbf{A}(t_{2})\cdot(\widehat{\mathbf{p}}-\widehat{\mathbf{Q}%
}(t_{2}))\right]  |\Phi_{\mathbf{k}}^{V}(\mathbf{r})\rangle
\nonumber\label{vpk-F}\\
&  \overset{\mathbf{q}=\mathbf{k-p}}{=}\frac{N_{p}^{\ast}}{\pi^{2}}%
\frac{\mathbf{A}\left(  t_{2}\right)  \cdot\mathbf{q}}{|\mathbf{q}|^{2}\left(
|\mathbf{q}|^{2}+2\mathbf{q}\cdot\mathbf{p}\right)  }\left(  \frac
{|\mathbf{q}|^{2}}{|\mathbf{q}|^{2}+2\mathbf{q}\cdot\mathbf{p}}\right)
^{i\eta_{p}}.
\end{align}
In ICSFA, they are
\end{subequations}
\begin{subequations}
\begin{align}
V_{\mathbf{k}g} &  =\langle\Phi_{\mathbf{k}}^{CV}(\mathbf{r})|\left[
\mathbf{A}(t_{1})\cdot\widehat{\mathbf{p}}+\frac{\mathbf{A}^{2}(t_{1})}{2}\right]
|\phi_{i}(\mathbf{r})\rangle\nonumber\label{vkg-I}\\
&  =(\frac{8}{\pi^{2}})^{1/2}N_{k}^{\ast}(1+i\eta_{k})\frac{\mathbf{A}\left(
t_{1}\right)  \cdot\mathbf{k}}{\left(  1+{k}^{2}\right)  ^{2}}\left(
\frac{1+ik}{1-ik}\right)  ^{i\eta_{k}},\\
V_{\mathbf{pk}} &  =\langle\Phi_{\mathbf{p}}^{V}\left(  \mathbf{r}\right)
|-\frac{1}{r}|\Phi_{\mathbf{k}}^{CV}(\mathbf{r})\rangle\nonumber\label{vpk-I}%
\\
&  \overset{\mathbf{q}=\mathbf{k}-\mathbf{p}}{=}-\frac{N_{k}}{2\pi^{2}}%
\frac{1}{|\mathbf{q}|^{2}}\left(  \frac{|\mathbf{q}|^{2}}{|\mathbf{q}%
|^{2}-2\mathbf{q}\cdot\mathbf{k}}\right)  ^{-i\eta_{k}}.
\end{align}
\end{subequations}

The prefactor $V_{\mathbf{k}g}$ affects the ionization yield of the electron
from the ground state to the continuum state with momentum $\mathbf{k}$.
$V_{\mathbf{pk}}$ determines the scattering cross-section in the elastic
scattering process of the freed electron from momentum $\mathbf{k}$ to
$\mathbf{p}$. In comparison to the standard SFA, FCSFA mainly corrects
$V_{\mathbf{pk}}$, while $V_{\mathbf{k}g}$ does not change. While in ICSFA,
both the $V_{\mathbf{k}g}$ and $V_{\mathbf{pk}}$ are different with that in
the SFA.

It is important to emphasize that, although we assumes a
pure Coulomb potential for simplicity in this derivation, the
ICSFA framework can be straightforwardly generalized to arbitrary
central or polycentric potentials describing more complex systems.
For multi-electron atoms, the potential $V(r)$ can be replaced by
a screened Coulomb potential that accounts for electron shielding,
or a static effective potential fitted to experimental data. For
molecular targets, the single-center Coulomb potential is extended
to a polycentric potential that describes the combined nuclear
attraction and electron distribution of multiple nuclei. Regarding
the prefactors in the ICSFA framework, for simple systems (e.g.,
atomic hydrogen), analytical expressions exist and maintain
computations tractable. However, for more complex targets as
multi-electron atoms or molecules, analytical expressions for
these prefactors are often unavailable, requiring numerical
integration--significantly increasing the computational cost.

The multi-dimensional integrations in Eq. (\ref{M}) are approximately solved
by the saddle-point method
%\end{subequations}
\cite{Kopold1999,Lewenstein1994,Lewenstein1995,Figueira2002}
\begin{align}
M_{resc}(\mathbf{p})  &  =\sum\limits_{j}(2\pi i)^{5/2}\frac{V_{\mathbf{pk}%
_{j}}V_{\mathbf{k}_{j}g}}{\sqrt{\det S_{\mathbf{p}}^{\prime\prime}(t_{1}%
,t_{2},\mathbf{k})\left\vert _{j}\right.  }}\nonumber\\
&  \times\exp\left[  iS_{\mathbf{p}}(t_{1j},t_{2j},\mathbf{k}_{j})\right]  ,
\end{align}
where the index $j$\ runs over the relevant saddle points, $S_{\mathbf{p}%
}^{\prime\prime}\left(  t_{1},t_{2},\mathbf{k}\right)  \left\vert _{j}\right.
$ denotes the five-dimensional matrix of the second derivatives to the action
in Eq. (\ref{S}) with respective to $t_{1},t_{2}$ and $\mathbf{k}$. The
corresponding saddle-point equations are\ \
\begin{equation}
\left[  \mathbf{k}+\mathbf{A}(t_{1})\right]  ^{2}=-2I_{p}, \label{equ:1}%
\end{equation}%
\begin{equation}
\left[  \mathbf{p}+\mathbf{A}(t_{2})\right]  ^{2}=\left[  \mathbf{k}%
+\mathbf{A}(t_{2})\right]  ^{2}, \label{equ:t}%
\end{equation}%
\begin{equation}
\int_{t_{1}}^{t_{2}}d\tau\left[  \mathbf{k}+\mathbf{A}(\tau)\right]  =0.
\label{equ:k}%
\end{equation}
These saddle-point equations are shared across the SFA, ICSFA, and FCSFA
frameworks. Eventually, $|M_{resc}(\mathbf{p})|^{2}$ is calculated for
different $\mathbf{p}$ to obtain the spectrum.

\begin{figure}[h]
\centering
\includegraphics[width=3in]{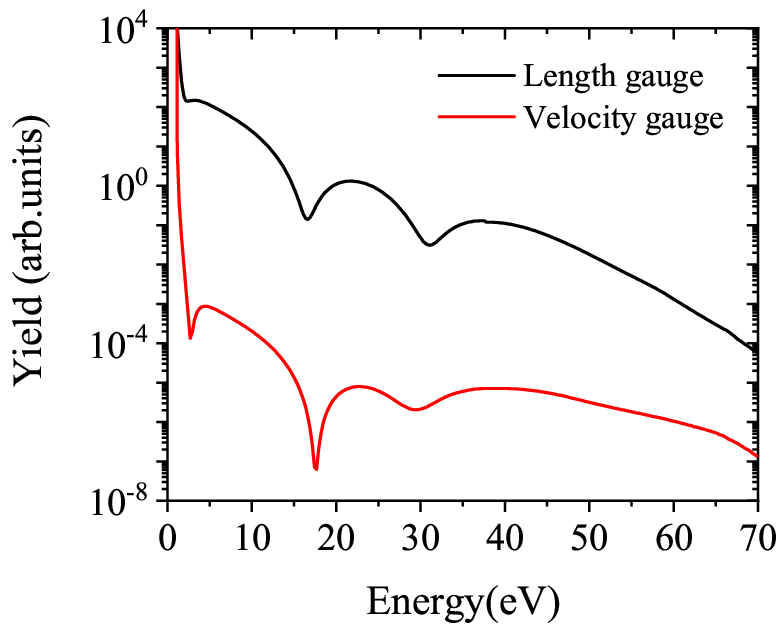}
\caption{(color online)
The photoelectron spectra of hydrogen atoms exposed in linearly polarized
laser fields ($I=1.0\times10^{14}$ W/cm$^{2}$ and $\lambda=800$ nm) calculated
in length gauge (black line) and velocity gauge (red line).}%
\label{figs1}%
\end{figure}

As is noted, the aforementioned formulas are derived within the velocity
gauge. In what follows, we proceed to present the ICSFA S-matrix series in the
length gauge. Specifically, during the derivation, one only needs to replace
the interaction term in Eq. (\ref{VI}) with $V_{I}=\mathbf{E}(t)\cdot
\mathbf{r}$ (where $\mathbf{E}\left(  t\right)  =-\frac{\partial
\mathbf{A}\left(  t\right)  }{\partial t}$), and substitute the wave functions
of Volkov and Coulomb-Volkov states with their length-gauge counterparts,
which are phase-shifted by $e^{i\mathbf{A}\left(  t\right)  \cdot\mathbf{r}}$
relative to those in the velocity-gauge formulation. Thereby, the ICSFA
S-matrix series in the length gauge can be obtained as follows (denoted by the
symbol L):

\begin{align}
S_{fi}^{I(1);L}=  &  -i\int dt_{1}\langle\Phi_{\widetilde{\mathbf{p}}}%
^{V}\left(  \mathbf{r}_{1},t_{1}\right)  |\nonumber\\
&  \times\left[  \mathbf{E}(t_{1})\cdot\mathbf{r}_{1}\right]  \left\vert
\phi_{i}\left(  \mathbf{r}_{1},t_{1}\right)  \right\rangle ,
\end{align}%
\begin{align}
S_{fi}^{I(2);L}=  &  -i\int\int_{t_{1}}dt_{2}dt_{1}\left\langle \Phi
_{\widetilde{\mathbf{p}}}^{V}\left(  \mathbf{r}_{2},t_{2}\right)  \right\vert
V(r_{2}) \nonumber\\
&  \times G_{CV}^{(L)}\left(  \mathbf{r}_{2},t_{2};\mathbf{r}_{1},t_{1}\right)
\nonumber\\
&  \times\left[  \mathbf{E}(t_{1})\cdot\mathbf{r}_{1}\right]  \left\vert
\phi_{i}\left(  \mathbf{r}_{1},t_{1}\right)  \right\rangle ,
\end{align}
and the nth-order term%
\begin{align}
S_{fi}^{I(n);L}=  &  -i\int\cdot\cdot\cdot\int_{t_{n-2}}\int_{t_{n-1}}%
dt_{n}dt_{n-1}\cdot\cdot\cdot dt_{1}\nonumber\\
&  \times\langle\Phi_{\widetilde{\mathbf{p}}}^{V}(\mathbf{r}_{n}%
,t_{n})|V(r_{n})G_{CV}^{(L)}(\mathbf{r}_{n},t_{n};\mathbf{r}_{n-1},t_{n-1}%
)\nonumber\\
&  \times  \left\{V(r_{n-1})+\frac{Z}{r_{n-1}}+\mathbf{A}(t_{n-1})\cdot\left[\widehat{\widetilde{\mathbf{p}}
}-\widehat{\mathbf{Q}}^{(L)}(t_{n-1})\right]\right\} \nonumber\\
&  \times\cdot\cdot\cdot\times\nonumber\\
&  \times G_{CV}^{(L)}(\mathbf{r}_{2},t;\mathbf{r}_{1},t_{1})\left[
\mathbf{E}(t_{1})\cdot\mathbf{r}_{1}\right] \nonumber\\
&  \times|\phi_{i}(\mathbf{r}_{1},t_{1})\rangle.
\end{align}
where, $\widetilde{\mathbf{p}}=\mathbf{p+A}\left(  t\right)  $, $\langle
\mathbf{r}|\Phi_{\widetilde{\mathbf{p}}}^{V}(t)\rangle=e^{i\mathbf{A}\left(
t\right)  \cdot\mathbf{r}}\langle\mathbf{r}|\Phi_{\mathbf{p}}^{V}(t)\rangle$,
$\langle\mathbf{r}|\phi_{\widetilde{\mathbf{p}}}^{C}\rangle=e^{i\mathbf{A}%
\left(  t\right)  \cdot\mathbf{r}}\langle\mathbf{r}|\phi_{\mathbf{p}}%
^{C}\rangle$, and $\widehat{\mathbf{Q}}^{(L)}(t)=\underset{\mathbf{p}}{\sum
}|\phi_{\widetilde{\mathbf{p}}}^{C}\rangle\mathbf{p}\langle\phi_{\widetilde
{\mathbf{p}}}^{C}|$.

\begin{subequations}
The second-order term in the ICSFA S-matrix series can also be expressed using
the same formula as in Eq. (\ref{M}). Notably, the exponential phase factor
$e^{iS_{\mathbf{p}}(t_{1},t_{2},\mathbf{k})}$ is gauge-invariant. Below, we
derive the prefactors in the length gauge:
\begin{align}
V_{\mathbf{k}g}^{L}  &  =\langle\Phi_{\widetilde{\mathbf{k}}}^{CV}%
(\mathbf{r})|\mathbf{E}(t_{1})\cdot\mathbf{r}|\phi_{i}(\mathbf{r}%
)\rangle\nonumber\\
&  =\langle\Phi_{\mathbf{k}}^{CV}(\mathbf{r})|e^{-i\mathbf{A}\left(
t_{1}\right)  \cdot\mathbf{r}}\mathbf{E}(t_{1})\cdot\mathbf{r}|\phi
_{i}(\mathbf{r})\rangle\nonumber\\
&  =2^{5/2}N_{k}^{\ast}\frac{(1-i\eta_{k})M^{i\eta_{k}}\mathbf{E}(t_{1}%
)\cdot\left[  \mathbf{k}I_{2}-\mathbf{A}(t_{1})I_{1}\right]  }{\left\{
1+[\mathbf{A}\left(  t_{1}\right)  \mathbf{+k}]^{2}\right\}  ^{3}},
\end{align}
where%
\begin{align}
I_{1}  &  =\left[  -(2i+\eta_{k})+(2\eta_{k}-i)M+(\eta_{k}-i)\frac{1+i\eta
_{k}}{1-i\eta_{k}}M^{2}\right]  ,\nonumber\\
I_{2}  &  =\left[  2i+\eta_{k}-(\eta_{k}-i)M\right]  ,\nonumber
\end{align}
with
\[
M=\frac{\left\{  1+\left[  \mathbf{A}\left(  t_{1}\right)  \text{ }%
+\text{\ }\mathbf{k}\right]  ^{2}\right\}  }{\left[  \mathbf{A}^{2}\left(
t_{1}\right)  -(k+i)^{2}\right]  }.
\]
For the length-gauge expression of $V_{\mathbf{pk}}^{L}$, we have
\begin{align}
V_{\mathbf{pk}}^{L}  &  =\langle\Phi_{\mathbf{p}}^{V}(\mathbf{r}%
)|e^{-i\mathbf{A}\left(  t_{2}\right)  \cdot\mathbf{r}}\left[  -\frac{1}%
{r}\right]  e^{i\mathbf{A}\left(  t_{2}\right)  \cdot\mathbf{r}}%
|\Phi_{\mathbf{k}}^{CV}(\mathbf{r})\rangle\nonumber\\
&  =\langle\Phi_{\mathbf{p}}^{V}(\mathbf{r})|-\frac{1}{r}|\Phi_{\mathbf{k}%
}^{CV}(\mathbf{r})\rangle\nonumber\\
&  =V_{\mathbf{pk}}.
\end{align}
Therefore, $V_{\mathbf{pk}}$ remains gauge-invariant while $V_{\mathbf{k}g}$
explicitly depends on the gauge choice. Since the exponential phase factor
$e^{iS_{\mathbf{p}}(t_{1},t_{2},\mathbf{k})}$ and the prefactor
$V_{\mathbf{pk}}$ dominate the signature of the spectrum in the saddle-point
method, the overall spectral structure is gauge-invariant
\cite{FHM2007,Faisal2007,WBecker2009}.
\end{subequations}
This is exemplified by the comparison of photoelectron
spectra for hydrogen atoms exposed to linearly polarized laser
fields under the length and velocity gauges, as shown in Fig.
\ref{figs1}. The overall spectral structure--including the
positions of dominant peaks--is consistent across both gauges.
Minor discrepancies in the order of magnitude do not affect the
physical conclusions. All subsequent results, calculated in the
velocity gauge, are thus based on a gauge-invariant foundation.

\section{RESULTS AND DISCUSSION}

In this section, we systematically investigate how intermediate-state Coulomb
interaction shapes the spectra for a hydrogen atom
exposed to a linearly polarized laser field. All  photoelectron energy spectrum (PES) are calculated with
the detection direction aligned parallel to the laser polarization axis. To validate the newly developed
ICSFA framework, we first conduct a detailed comparative analysis of the three
S-matrix formulations against numerical solutions of the time-dependent
Schr\"{o}dinger equation (TDSE) calculations, a rigorous numerical benchmark.
The TDSE calculations were performed using QPROP \cite{Bauer2006}, a widely
recognized and freely available software package whose validity has been
rigorously validated in numerous prior studies.

\begin{figure*}[t]
\centering
\includegraphics[width=0.85\textwidth]{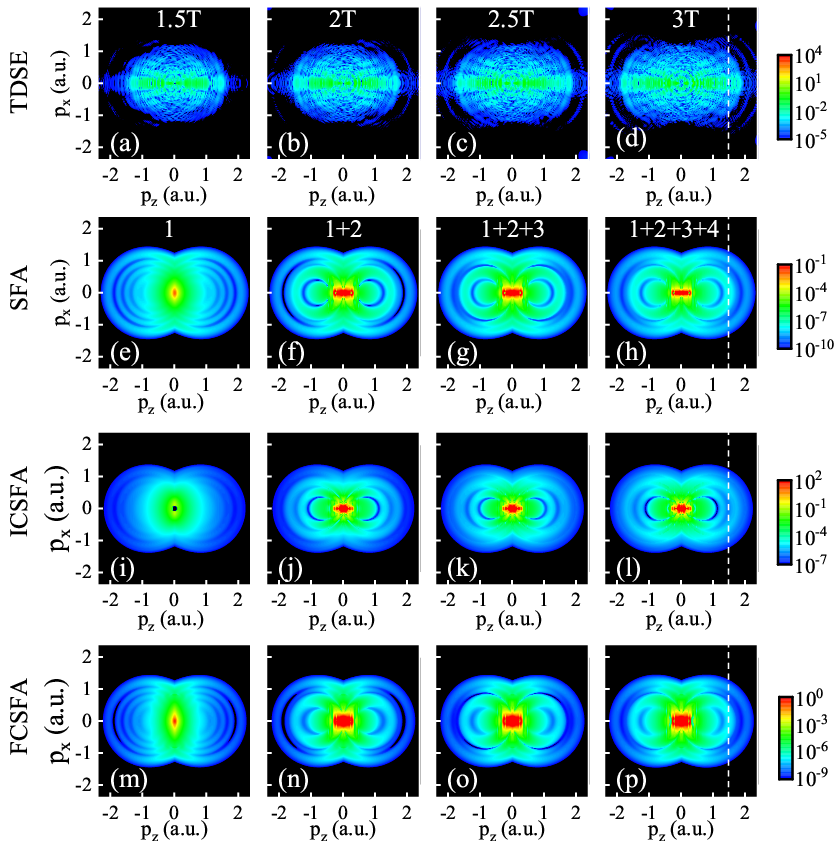}
\caption{(color online)
The logarithm of the ionization probability for H in an linearly polarized laser
with different pulse duration for TDSE method and different recollision trajectories
for S-matrix methods as depicted in the plots.
[(a)-(d)]: Spectra obtained by numerical solutions of the TDSE.
[(e)-(h)]: Spectra obtained with the SFA.
[(i)-(l)]: Spectra obtained with the ICSFA.
[(m)-(p)]: Spectra obtained with the FCSFA.
The laser parameters are $I=1.0\times10^{14}$ W/cm$^{2}$ and $\lambda=800$ nm.}%
\label{fig11}%
\end{figure*}

For the TDSE calculations, the linearly polarized laser field is described by
the vector potential: $\mathbf{A}\left(  t\right)  =A_{0}f(t)\cos(\omega
t)\widehat{\mathbf{e}}_{z}$, where $A_{0}$ is the amplitude, $f(t)$ is the
pulse envelope, $\widehat{\mathbf{e}}_{z}$ is the unit vector along the z-axis
(defining the polarization direction), and $\omega$ is the laser field
frequency. The amplitude $A_{0}$ of the laser field, expressed in atomic units,
is related to the intensity $I$ (W/cm$^{2}$) by $I=(A_{0}\omega)^2I_{A}$,
where $I_{A}=3.509\times10^{16}$ W/cm$^{2}$ denotes the atomic unit of intensity.
The ponderomotive energy of the electron in the laser field is given by $U_{p}=A_{0}^{2}/4$.
Here, we set $A_{0}=0.938$ a.u. and $\omega = 0.05695$ a.u.,
corresponding to a laser intensity of $I=$ $1.0\times10^{14}$ W/cm$^{2}$ and a
wavelength of 800 nm. To facilitate comparison with S-matrix theory, the pulse
envelope $f(t)$ is configured to consist of $(N+4)$ cycles (two-cycles
ramp-on, $N$-cycles of constant amplitude, and two-cycles ramp-off). $N$
cycles varied from 1.5 cycles to 3 cycles in increment of 0.5 cycle. This
pulse design ensures that only specific recollision trajectories contribute to
the spectrum for a given $N$. In the recollision process, the different
return-recollision trajectories are distinguished according to the travel time
$t_{t}$ defined as the interval between the ionization time and the
recollision time. For trajectories with $t_{t}$ in the interval
\{$(n/2)T,~[(n+1)/2]T$\} ($T$ is the optical cycle), we denote them as the
$n$th-return-recollision trajectories \cite{Hao2011}, which approach the
origin $n$ times, yet recollision occurs exclusively during the final return.
Therefore, only the first return trajectory occurs when $N=1.5$, with an
additional return trajectory added sequentially for every half-cycle increment
(e.g., for $N=2$, both the first and second returns are present, and so on).
This design allows clear examination of the contributions from different
return trajectories. Correspondingly, in the S-matrix calculations, we adopt
the same laser parameters as in the TDSE but employ a constant-amplitude pulse
(excluding ramp-on and ramp-off parts). Critically, there is a direct mapping
between the recollision trajectories included in S-matrix calculations and
TDSE pulse durations: limiting the calculation to the first return trajectory
corresponds to the $N=1.5$ pulse case in TDSE; including both first and second
returns aligns with $N=2$, and so forth for higher-order returns.

\begin{figure}[b]
\centering
\includegraphics[width=3in]{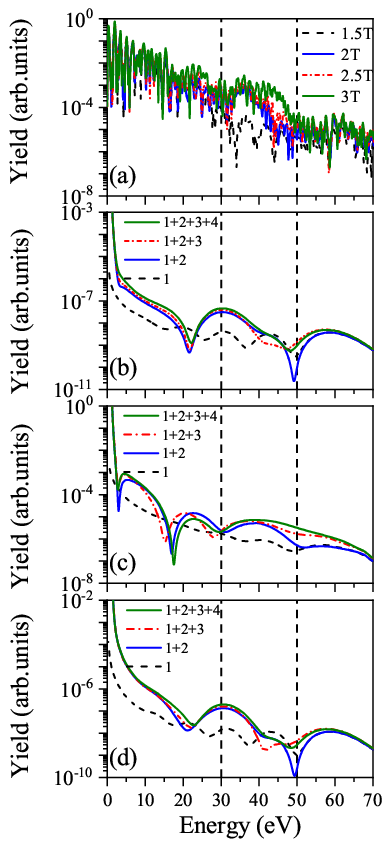}
\caption{(color online)
Photoelectron energy spectra (PES) of hydrogen atoms exposed in linearly
polarized laser fields ($I=1.0\times10^{14}$ W/cm$^{2}$, $\lambda=800$ nm)
calculated using different methods, compared with the TDSE results. Panels
(a)-(d) correspond to calculations using the TDSE, SFA, ICSFA and FCSFA,
respectively. All PES are calculated with the detection direction aligned
parallel to the laser polarization axis.}%
\label{fig0}%
\end{figure}

In figure \ref{fig11} we display the PMDs
calculated using the TDSE, SFA, ICSFA and FCSFA for
$I=1.0\times10^{14}$ W/cm$^{2}$, wavelength $\lambda=800$ nm.
While TDSE simulations capture all ionization pathways (direct
ionization and recollision-mediated ionization), the second-order
S-matrix series solely accounts for recollision process. We
therefore focus on the high-momentum region, where recollision
process dominates, across all methods. We compare the PMDs
calculated by TDSE for laser pulse durations ranging from 1.5T to
3T with those obtained using the aforementioned methods, which
includes contributions from different return-recollision
trajectories, with the pulse duration and the number of the
return-recollision trajectories indicated in the corresponding
plots. The notations: "1" denotes only the first return; "1+2"
adds the second return; "1+2+3" further incorporates trajectories
up to the third return; and "1+2+3+4" includes all contributions
from the first- to the fourth returns. The top row displays the
results obtained with TDSE, followed sequentially in the next
three rows by the results calculated using the SFA, ICSFA, and
FCSFA methods.

All distributions exhibit an approximately
ellipsoidal shape. For TDSE results, as the pulse duration
increases (from left to right), the PMD gradually expands outward
(to higher momenta). This trend directly demonstrates the crucial
role of the MRR trajectories in shaping the final momentum
distribution, particularly in the high-momentum region A similar
trend is observed in ICSFA (Figs. \ref{fig11}(i-l)): the
ionization probability in the high-momentum region gradually
increases as multi-return trajectories are added incrementally. In
contrast, results from SFA (Figs. \ref{fig11}(e-h)) and FCSFA
(Figs. \ref{fig11}(m-p)) show significant outward expansion upon
adding the second return, but higher-order returns (3rd/4th)
induce negligible changes.

Clear ring structures are also present in the SFA,
ICSFA, and FCSFA results. Critically, these are not ATI rings but
arise from interference between recollision trajectories,
including interference between long and short trajectories within
individual returns and interference among different returns.
Notably, SFA and FCSFA exhibits nearly identical ring positions,
whereas ICSFA ring positions differ. For example, at $p_{z}
\approx 1.485$ a.u. (indicated by the vertical dashed lines in the
figures), ICSFA shows a dip (destructive interference), while SFA
and FCSFA exhibit a peak (constructive interference).

\begin{figure}[b]
\centering
\includegraphics[width=3in]{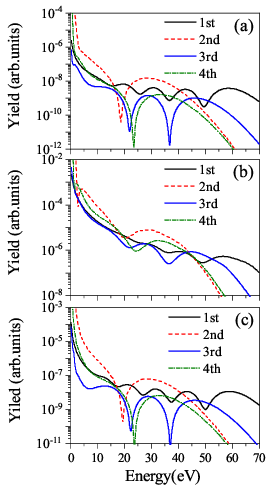}
\caption{(color online)
Separate contributions of different recollision trajectories calculated
using the three methods: (a) SFA, (b) ICSFA, and (c) FCSFA. The laser
parameters are identical to those employed in Fig. \ref{fig0}.}%
\label{fig2}%
\end{figure}

To quantitatively compare the results of the three
methods and investigate the structural features in the high-energy
region, we perform further analysis using the PES along the laser
polarization direction. Figure \ref{fig0} displays the PES for H
atoms ionized in $1.0\times10^{14}$ W/cm$^{2}$ laser field,
calculated by different theoretical frameworks.
 For TDSE-simulated spectra (Fig. \ref{fig0}(a)), we observe a
significant rise in intensity within 30-50 eV (marked by dashed lines) as the
pulse duration increases from 1.5 to 3 optical cycles. Among the three
S-matrix frameworks, it can be observed that the structures in the PES
calculated with SFA (Fig. \ref{fig0}(b)) and FCSFA (Fig. \ref{fig0}(d)) are
essentially identical, despite notable discrepancies in yield. In comparison,
the spectrum calculated with ICSFA (Fig. \ref{fig0}(c)) exhibits distinct
structures from those calculated with SFA or FCSFA. As the pulse duration
increases, both SFA and FCSFA predict yield enhancements when incorporating
additional trajectories. However, the energy regions exhibiting these
enhancements show significant deviations from TDSE benchmarks. In contrast,
the ICSFA-simulated spectrum maintains spectral features consistent with TDSE
results across the same energy range under increasing pulse durations. This
agreement directly validates the ICSFA framework while highlighting the
essential contribution of MRR trajectories to the spectrum in high-energy region.

Next, we performed a detailed analysis of the contributions of different
return-recollision trajectories ranging from the first- to the fourth-return
to the PES obtained using the three S-matrix frameworks, as shown in Fig.
\ref{fig2}. Among the results calculated by all methods, the cutoff energy
corresponding to the spectra of the first-return trajectories is the highest.
This is followed in order by the third-, fourth- and second-return
trajectories, which is consistent with the usual predictions of simple-man
theory. However, the relative contribution of each return obtained by the
different methods varies. In both SFA and FCSFA cases, as demonstrated in
Figs. \ref{fig2}(a) and \ref{fig2}(c), the first- and second-return
trajectories dominate the PES. Specifically, the second return dominates the
PES in the low- and medium-energy regions (i.e., energies less than 40 eV),
while the first return dominates in the high-energy region (i.e., energies
greater than 40 eV). The contribution of the third- and fourth-return
trajectories is consistently smaller than that of the first- and second-return
trajectories. This is because higher-order returns are associated with a
reduced probability of recollision with the ion due to the expansion of the
wave packet. Nevertheless, in the case of ICSFA in Fig. \ref{fig2}(b), the
contributions of the third and fourth returns are considerably higher. At
25-40 eV, the fourth return contributes more than the first return, and at
40-50 eV, the contribution of the third return exceeds that of the first
return. The observed increase in the contributions of MRR trajectories,
attributable to the intermediate-state Coulomb interaction, in this study is
consistent with the findings reported in Ref. \cite{Hao2022}.

\begin{figure}[t]
\centering
\includegraphics[width=3.5in]{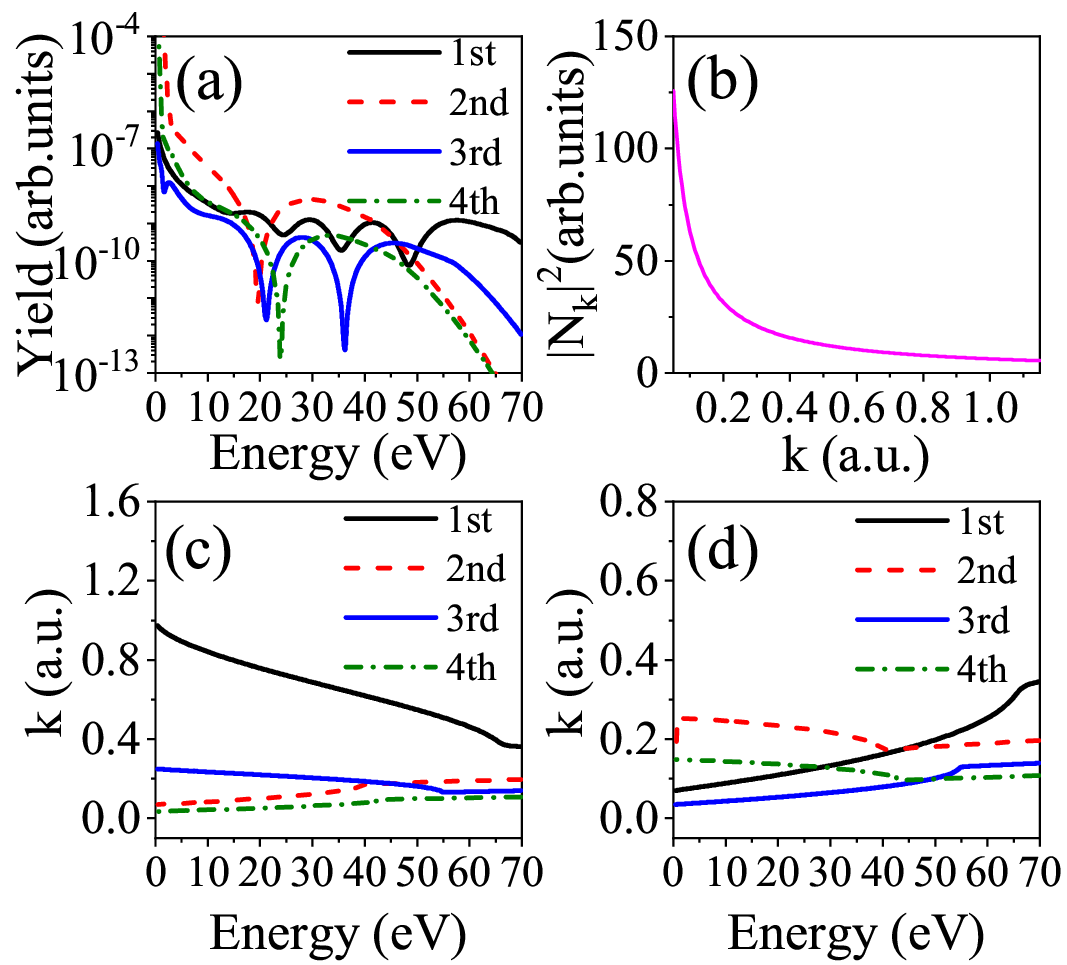} \caption{(color online) (a) shows the
contributions from the first- to the fourth-return-recollision trajectories,
calculated using the ICSFA method, without accounting for the $N_{k}$ term in
Eqs. (\ref{vkg-I}) and (\ref{vpk-I}). (b) depicts the variation of $N_{k}$ as
a function of photoelectron momentum $k$. (c) and (d) illustrate the
dependence of $k$ on the photoelectron energy for the short and long orbits in
each return, respectively. The laser parameters adopted herein are consistent
with those in Fig. \ref{fig0}.}%
\label{fig6}%
\end{figure}

\begin{figure}[t]
\centering
\includegraphics[width=3in]{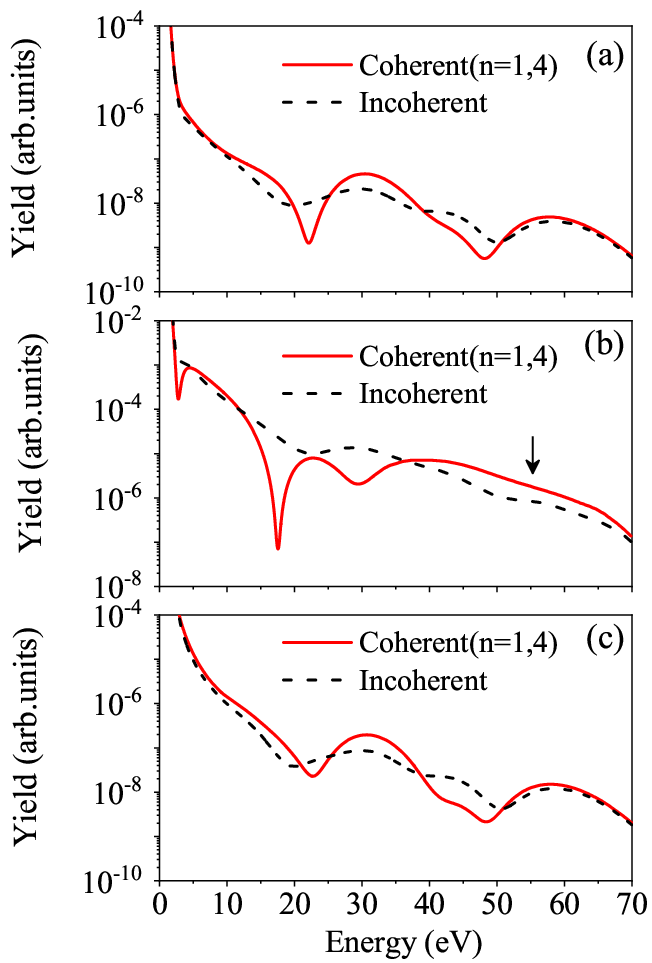}
\caption{(color online)
Photoelectron energy spectra calculated via (a) SFA, (b) ICSFA, and (c) FCSFA, respectively,
arising from the coherent or incoherent superposition of distinct return-recollision
trajectories (from the first return to the fourth return).
The coherent results are reproduced from Fig. \ref{fig0}.}%
\label{fig3}%
\end{figure}

As illustrated in the depiction of the methods, the saddle-point equations
(Eqs. (\ref{equ:1})-(\ref{equ:k})) and the solutions of the trajectories are
equivalent across the three methods. The distinction among the three methods
is rooted in the prefactors $V_{\mathbf{k}g}$ and $V_{\mathbf{pk}}$
articulated in Eqs. (\ref{vkg})-(\ref{vpk-I}). A thorough examination has
revealed that the normalization coefficient $N_{k}$ in $V_{\mathbf{k}g}$
(Eq. (\ref{vkg-I}))and $V_{\mathbf{pk}}$ (Eq. (\ref{vpk-I})) of the ICSFA
formula is instrumental in the reshaping of the relative contributions
associated with different return-recollision trajectories. As illustrated in
Fig. \ref{fig6}(a), when this coefficient is excluded from the ICSFA method,
the relative contributions of the third and fourth returns are substantially
diminished compared to Fig. \ref{fig2}(b), where $N_{k}$ is incorporated. The
normalization coefficient of the Coulomb-Volkov wave function of the
intermediate state, $N_{k}$, is expressed in Eq. (\ref{Np}). It has been
demonstrated that $N_{k}$ is contingent on the intermediate canonical momentum
$k$, exhibiting a decrease in value as $k$ increases, as illustrated in Fig.
\ref{fig6}(b). In fact, the value of $k$ varies with different returns. For
each return, there is a pair of recollision trajectories, designated as the
long and short orbits, which also have different $k$. The dependence of $k$ on
the photoelectron energy for the short and long orbits in each return is
presented in Figs. \ref{fig6}(c) and \ref{fig6}(d), respectively. It has been
observed that for the short orbits depicted in Fig. \ref{fig6}(c), the
intermediate canonical momentum $k$ for the first return is the greatest and
is considerably higher than the other returns, while the fourth return is the
lowest. For long orbits in Fig. \ref{fig6}(d), at energies less than 40 eV,
the second return is the highest, while at energies greater than 40 eV, the
first return is the highest. A comparative analysis reveals that the $k$
values for the third- and fourth-return trajectories are, on average,
considerably smaller than those for the first and second returns. As
demonstrated in Fig. \ref{fig6}(b), the values of $N_{k}$ for the third and fourth
returns are greater than those for the first and second returns, as a result
of the dependence of $N_{k}$ on $k$. This phenomenon results in a
corresponding increase in the prefactors $V_{\mathbf{k}g}$ and $V_{\mathbf{pk}%
}$, both of which are proportional to $N_{k}$. This, in turn, leads to an
increase in the relative contributions for the third and fourth returns. It is
evident that the ionization yield and the scattering cross-section of the
ionization process are contingent on $V_{\mathbf{k}g}$ and $V_{\mathbf{pk}}$,
respectively. Consequently, the intermediate-state Coulomb interaction has
been shown to enhance the third and fourth return trajectories by modifying
the ionization yield \cite{Faisal2018} and the scattering cross-section
\cite{Landau1965}.

It is important to note that the effect of
intermediate-state Coulomb interaction observed in this study is
closely linked to the Coulomb focusing effect in the semiclassical
picture. A smaller intermediate canonical momentum corresponds to
reduced effective kinetic energy, rendering the electron more
vulnerable to the attractive Coulomb potential--so much so that
the Coulomb field can more readily bend its trajectory toward the
parent ion. This enhancement of the electron wave packet's overlap
with the ion directly increases the density of intermediate
scattered states quantified by $N_{k}$. The Coulomb focusing
effect thus elevates the probability of multiple returns, a trend
precisely reflected in the larger $N_{k}$ for MRR trajectories and
ultimately explaining their preferential enhancement.

\begin{figure}[t]
\centering
\includegraphics[width=3.4in]{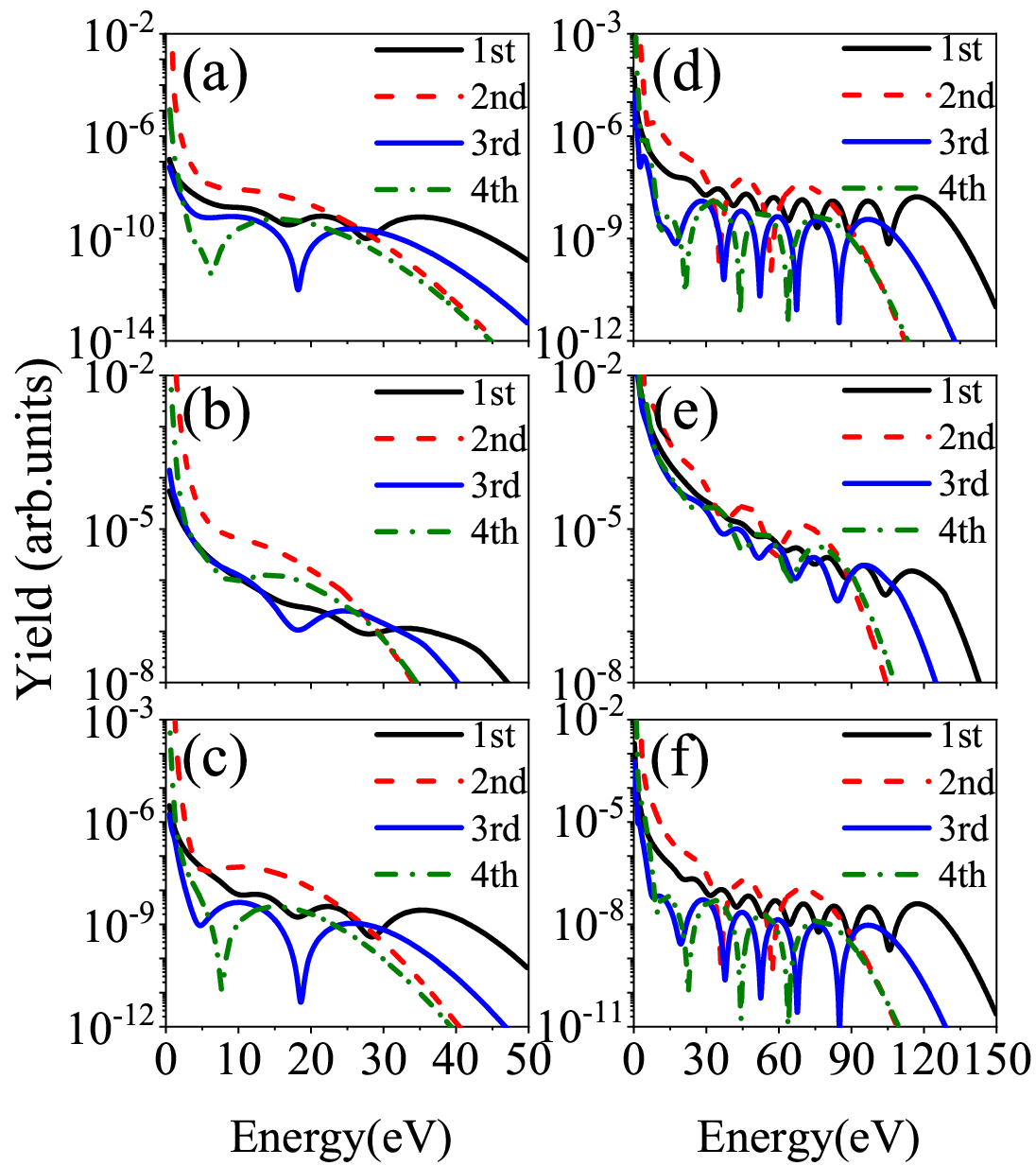} \caption{(color online) Photoelectron
energy spectra at 800 nm corresponding to different recollision trajectories,
calculated using the three methods: (a) and (d) SFA, (b) and (e) ICSFA, (c)
and (f) FCSFA. Laser intensities: $I=0.6\times10^{14}$ W/cm$^{2}$ for (a)-(c)
and $I=2.0\times10^{14}$ W/cm$^{2}$ for (d)-(f).}%
\label{fig4}%
\end{figure}

Changes in the relative contributions of different recollision
trajectories will also influence the outcomes of their interference. This is
clearly demonstrated in Fig. \ref{fig3} which presents the simulated spectra
resulting from incoherent and coherent superposition on recollision
trajectories with different returns for comparison. Adding recollision
trajectories with varying returns incoherently results in a more uniform,
structureless spectral profile. According to Fig. \ref{fig2}, the
intermediate-state Coulomb correction improves the relative contribution of
the third- and fourth-return trajectories. This effectively changes the
interference results between recollision trajectories with different returns;
this is why the interference structures in the PES obtained by ICSFA in Fig.
\ref{fig2} differ from those obtained by SFA and FCSFA. Furthermore, closer
inspection of Fig. \ref{fig3} reveals that for SFA and FCSFA,
no observable discrepancy exists between incoherent and
coherent results at energies higher than 50 eV. However, in the
case of ICSFA, a clear interference-induced enhancement
can still be observed in the high energy region, as indicated by
the arrow in Fig. \ref{fig3}(b). This is because the difference
between the contributions from the first and third returns in the
high-energy region is too great to produce an interference
structure in the results calculated by SFA and FCSFA. In the case
of ICSFA, however, there is a significant reduction in the gap
between the contributions from the first and third returns,
resulting in constructive interference.

\begin{figure}[t]
\centering
\includegraphics[width=3.3in]{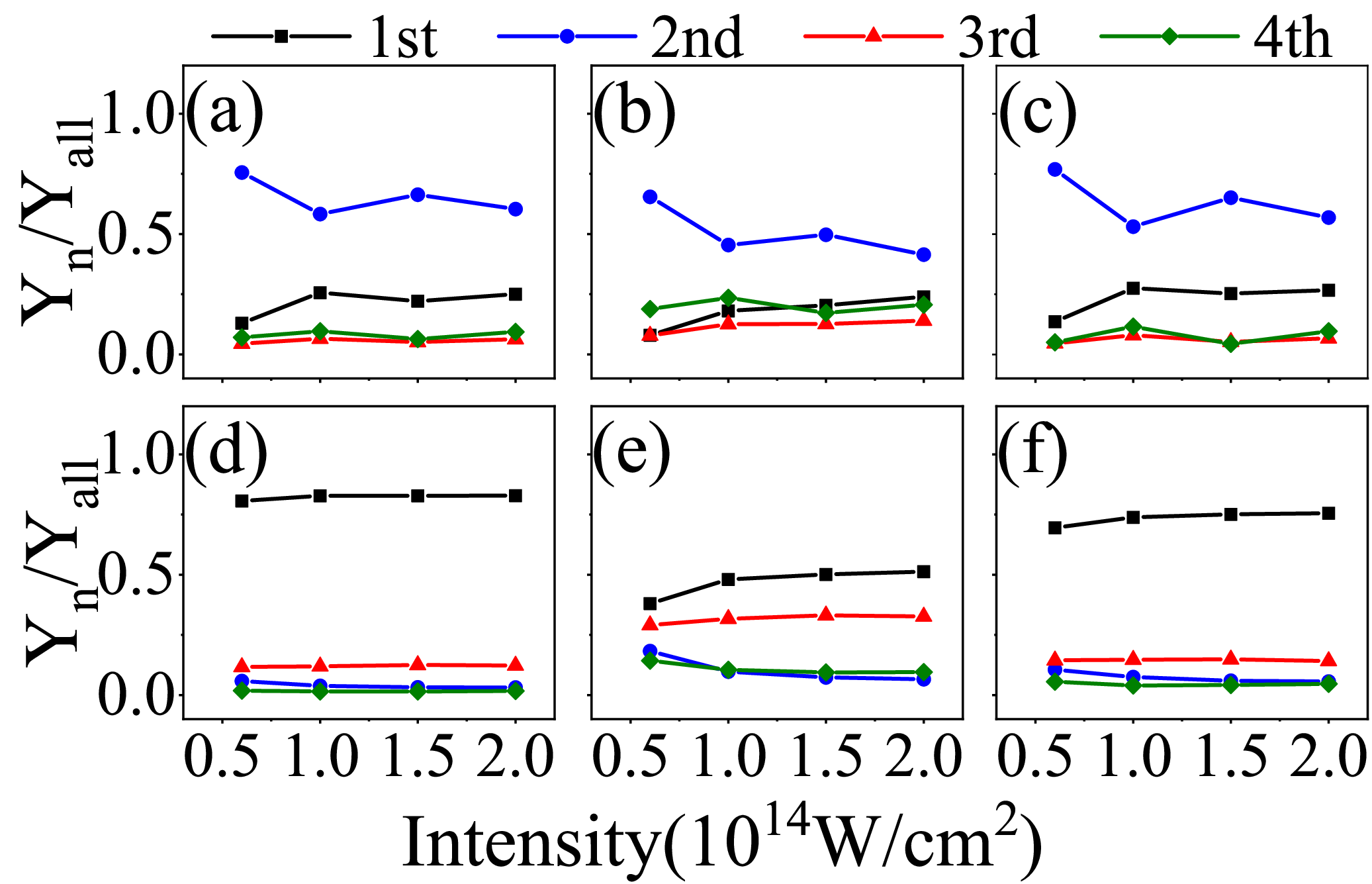} \caption{(color online) Dependence of
the proportion of different recollision trajectories on laser intensity within
a specific energy range. Upper panels: ratios of the integrated yield for each
return to the total yield in the $2U_{p}-7U_{p}$ energy range. Lower panels:
results calculated for energies greater than $7 U_{p}$. (a)/(d), (b)/(e), and
(c)/(f) correspond to calculations using SFA, ICSFA, and FCSFA, respectively.}%
\label{fig5}%
\end{figure}

We also calculated the PES at lower laser intensity of
$0.6\times10^{14}$ W/cm$^{2}$ and higher laser intensity of $2.0\times10^{14}$
W/cm$^{2}$, as shown in Fig. \ref{fig4}. We found that the conclusions are
similar under different laser intensities. That is, the contributions of the
third- and fourth-return trajectories calculated by SFA and FCSFA methods are
significantly smaller than those of the first and second returns. While in the
case of ICSFA, the contribution of the third- and fourth-return trajectories
significantly increased, even exceeding the contribution of the first and
second returns at specific energies.

In order to quantify the impact of intermediate-state Coulomb interaction on
the relative contributions of different recollision trajectories, the
proportion of different recollision trajectories is presented in Fig.
\ref{fig5}. This proportion is defined as the ratio of the integrated yield
for each return to the total yield of all returns in a specific energy region.
The present study focuses on two distinct energy regions: the intermediate
energy region, which is defined as the range of $2 U_{p}<E<7 U_{p}$, and the
high energy region, which is defined as the range of $E>7 U_{p}$. In the
intermediate energy region (the upper panels in Fig. \ref{fig5}), the
proportion for the second return is found to be predominant over the other
returns in the results obtained by all three methods. Nevertheless, a
comparative analysis of the other returns reveals divergent behaviors in the
results obtained with various methods. As indicated by the results of SFA and
FCSFA in Fig. \ref{fig5}(a) and \ref{fig5}(c), the proportion of the first
return is notably higher than those of the third and fourth returns at
different laser intensities. However, in the case of ICSFA, both the
contributions of the third and fourth returns are significantly improved,
reaching a level comparable to that of the first return. This phenomenon is
particularly evident at lower laser intensities, where the proportion of the
fourth return exceeds that of the first return. In the high-energy region
(lower panels in Fig. \ref{fig5}), the proportion for the first return is
notably predominant over the other returns in the results obtained by all
three methods. The primary distinction manifests specifically in the third
return. The results of SFA and FCSFA demonstrate that the contribution of the
third return is significantly lower than that of the first return at all laser
intensities considered in this study. However, the results of ICSFA
demonstrate that the intermediate-state Coulomb interaction significantly
increases the contribution of the third return at all laser intensities. This
increase renders the third return comparable to the first return in the high
energy region.

\section{CONCLUSION}

In this paper, we obtain a complete S-matrix series that incorporates the
intermediate-state Coulomb interaction (ICSFA) in all orders. A systematic
investigation was conducted to determine the influence of the
intermediate-state Coulomb interaction on the PMDs and the energy spectra of hydrogen atoms
exposed to a linearly polarized laser field. This
investigation involved the application to the second-order term, which
describes the rescattering process. We present a comparative analysis of
photoelectron energy spectra computed via the SFA, ICSFA, and FCSFA
approaches, benchmarked against TDSE simulations. The ICSFA-simulated results
exhibit substantially better agreement with TDSE calculations compared to
those from SFA and FCSFA. Intermediate-state Coulomb corrections are found to
induce a significant enhancement of the relative contributions of the third-
and fourth-return-recollision trajectories. This observation is consistent
with the findings of previous joint theoretical and experimental studies. This
enhancement results in the presence of distinct interference structures in the
energy spectra compared with SFA and FCSFA. The enhancement is attributed to
intermediate-state Coulomb effects, which modify both the ionization yield and
the scattering cross-section via the Coulomb normalization factor, $N_{k}$.
These effects are equivalent to the so-called Coulomb focusing effect. The
intermediate-state Coulomb effects are found to persist across a broad
intensity range. The developed ICSFA formalism provides a foundation for
future investigations of strong-field processes based recollision, with direct
implications for high-order harmonic generation and non-sequential double
ionization for atoms and molecules.

\section{ACKNOWLEDGMENT}

Project supported by the National Natural Science Foundation of China (Grant
Nos. 12274273, 12450402), Innovation Program for Quantum Science and
Technology (No. 2021ZD0302101).

\end{document}